# Number π from the decoration of the Langstrup plate

**Amelia Carolina Sparavigna**
Department of Applied Science and Technology
Politecnico di Torino, Italy

*Studies of ancient bronze artifacts can be useful in understanding the progression of human knowledge of mathematics and geometry. Here I discuss the decoration composed by several circles and spirals of the Langstrup belt disk, an artifact of the Bronze Age found in Denmark. I am showing by measurements of diameters and distances of spirals, that the artist who made the decoration knew some approximations by rational numbers of the number π, the dimensionless physical quantity representing the ratio of circumference to diameter.*

In 2006, the Danish National Bank proposed a new series of banknotes. Karin Birgitte Lund, the artist who developed the design of the series, decided to represent some objects of the Bronze Age found in Denmark. Among them there are the Trundholm sun chariot and the Langstrup belt plate (see Fig.1, [1,2]).

I have already discussed the Trundholm sun chariot in a previous paper [3], where I proposed a calendar of 360 days based on it. In [3], I concluded telling that further studies of the decorations of ancient bronze artifacts can be useful to understand the progression of human knowledge of mathematics and geometry. In fact, the other previously mentioned ancient object, the Langstrup belt, can help us in the investigation of the knowledge of number π, the dimensionless physical quantity, which is the ratio of any Euclidean circle's circumference to its diameter.

The Langstrup belt plate was found before 1880 together with a bronze knife and spiral bangles. It is coming from the early Bronze Age, approximately 1400 BC. The decoration is composed by circular grooves and spirals (see-Fig.1), stamped probably by means of some standard punches into wax model before casting. Belt plates were worn by women on the front of their belts, as shown by the mummy of the "Egtved girl" [4].

In Ref.5, we find some interesting discussion on the figures emerging from the decoration. Let me report what this reference is telling about the Langstrup belt plate..."It has, apart from the point, four zones with 15+22+26+32 = 95 spirals in all. Still, a numerical pattern does not seem to emerge. However, if one … multiplies by the number of the factor of the zones, the sum of the spirals turns out to be 15×1+22×2+26×3+32×4 = 265, or exactly the number of days in 9 months of the Moon-year (265½), or, incidentally, also the length of the average human period of pregnancy. ... Going one step further, and again multiplying with the zonal factors, but now incorporating the point of the Langstrup belt-plate as Factor 1 (but with the value of 0), a sum of 0×1+15×2+22×3+26×4+32×5 = 360 appears."

Instead of using the numbers for a calendar, here I want to make some figures to investigate the knowledge of π, the ratio of circumference to diameter. As we will see in the following discussion, a numerical pattern is clearly emerging, because the artist prepared the decoration on the wax using π approximated by rational numbers, that is, fractions having integers in numerators and denominators.

Table 1 is proposing some data on the spirals: the number of spiral in each annulus (the region lying between two large concentric circles) and the number of turns in each spiral. For what concerns the turns, their numbers depend on the manner we count them. In the Figure 2 I show two possible numbers, according with two different radial directions on the spiral.

**Table 1** (for a check of the numbers see the Figures 1 and 2)

| Annulus | I | II | III | IV |
|---|---|---|---|---|
| N of spirals | 15 | 22 | 26 | 32 |
| N of spiral turns | 5 or 6 | 7 or 8 | 9 or 10 | 10 or 11 |

We can calculate the length L of the circumference which is the locus of the centres of spirals, to be two times the radius ($R_i$) of the spirals, multiplied by the number of spirals (see the second line of Table 2).

The radius of the spirals could be estimated by the spiral turns. Let us note that for the first and the second annuli, we see that there is a certain distance between the spirals. We could assume this distance as two times the thickness of a spiral turn. Therefore, we could calculate the lengths, using a number of turns of 7 instead of 6, for the first annulus and 9 for the second.

**Table 2** Length of the circumferences

| Annulus | I | II | III | IV |
|---|---|---|---|---|
| L | 15×2×$R_I$ | 22×2×$R_{II}$ | 26×2×$R_{III}$ | 32×2×$R_{IV}$ |
| L | 15×2×7 | 22×2×9 | 26×2×10 | 32×2×11 |

To have a possible evaluation of π, we need to give the radius of this circumference too. I decided the following approach to estimate the radius. I supposed that the artist used a multiple value of the spiral radius, measured on the radial direction, to determine the radius of the circumference suitable for a decoration with the given spiral. Therefore I found it, according to the Figure 3. The values of the radius are given in the following Table 3:

**Table 3**. Radius of the circumference of Table 2 (to estimate the radii see Figure 3)

| Annulus | I | II | III | IV |
|---|---|---|---|---|
| R | 5×$R_I$ | 7×$R_{II}$ | 8×$R_{III}$ | 10×$R_{IV}$ |

Using Table 2 and 3, we can then determine the value of π = L/(2R). Here the results in Table 4.

**Table 4**: π=L/(2R) as a rational number

| Annulus | I | II | III | IV |
|---|---|---|---|---|
| π | 15/5=3.0 | 22/7=3.1428 | 26/8=3.25 | 32/10=3.20 |

Supposing that the artist used for the radius a multiple of the spiral radius, Table 4 shown that the artist knew some approximation of number π. Let us remember that the disk is coming from the early Bronze Age, approximately 1400 BC.

There is also another possibility to evaluate π, that of calculate the ratio between the number of spirals and the number of turns in each spiral. Choosing for this number the lowest value, we have therefore the following Table 5.

**Table 5**

| Annulus | I | II | III | IV |
|---|---|---|---|---|
| N of spirals | 15 | 22 | 26 | 32 |
| N of spiral turns | 5 | 7 | 9 | 10 |
| π | 15/5=3.0 | 22/7=3.1428 | 26/9=2.89 | 32/10=3.20 |

In my opinion, **Table 4**, which is based on the measurements of radii and circumferences is the more realistic, showing clearly the origin of the number π as a dimensionless physical quantity.

In conclusion, once again, we can tell that these artefacts are fundamental to understand the knowledge of mathematics of ancient people. Besides being amazing, the decoration of the Langstrup belt disk demonstrates that who made it knew the number π as a ratio of integers.

**References**
1. http://en.wikipedia.org/wiki/Danish_krone
2. Bæltepladen fra Langstrup, http://natmus.dk/en/historisk-viden/danmark/moeder-med-danmarks-oldtid/the-bronze-age/baeltepladen-fra-langstrup/
http://www.jadu.de/mittelalter/germanen/gk/pages/guertelscheibe_jpg.htm
3. Amelia Carolina Sparavigna, Ancient bronze disks, decorations and calendars, 2012, arXiv:1203.2512v1 [physics.pop-ph], http://arxiv.org/abs/1203.2512
4.. Egtved Girl, http://en.wikipedia.org/wiki/Egtved_Girl
5. 2. Klavs Randsborg, SPIRALS! Calendars in the Bronze Age in Denmark, 2010, Adoranten. Vol.2009, http://www.ssfpa.se/pdf/2009/Randsborg.pdf

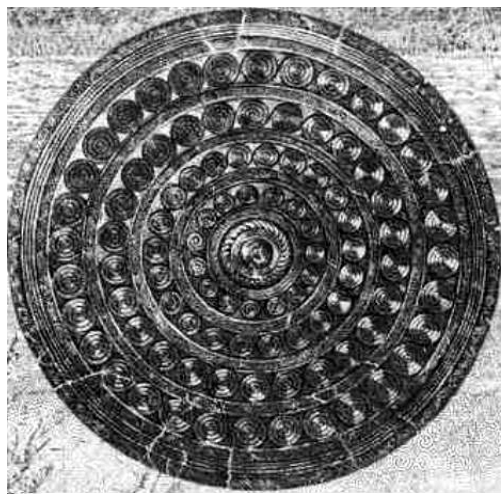

**Fig.1 The Langstrup belt plate on the Danish banknotes. We see four annuli, decorated with spirals**.

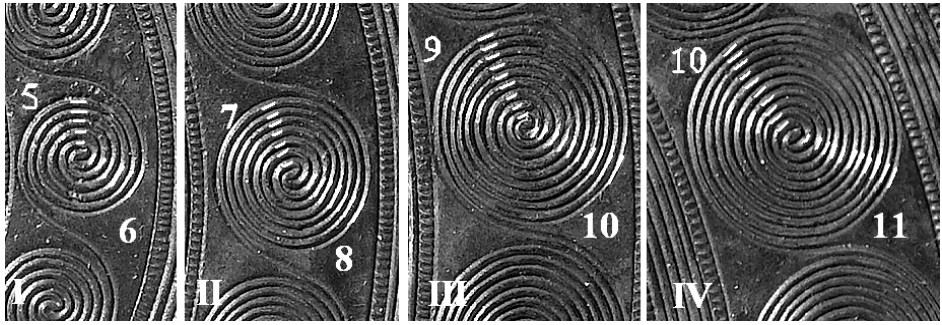

**Fig.2** A detail of the decoration. The numbers are the turns of the spiral, counted at two different radial directions. The roman number gives the corresponding annulus.

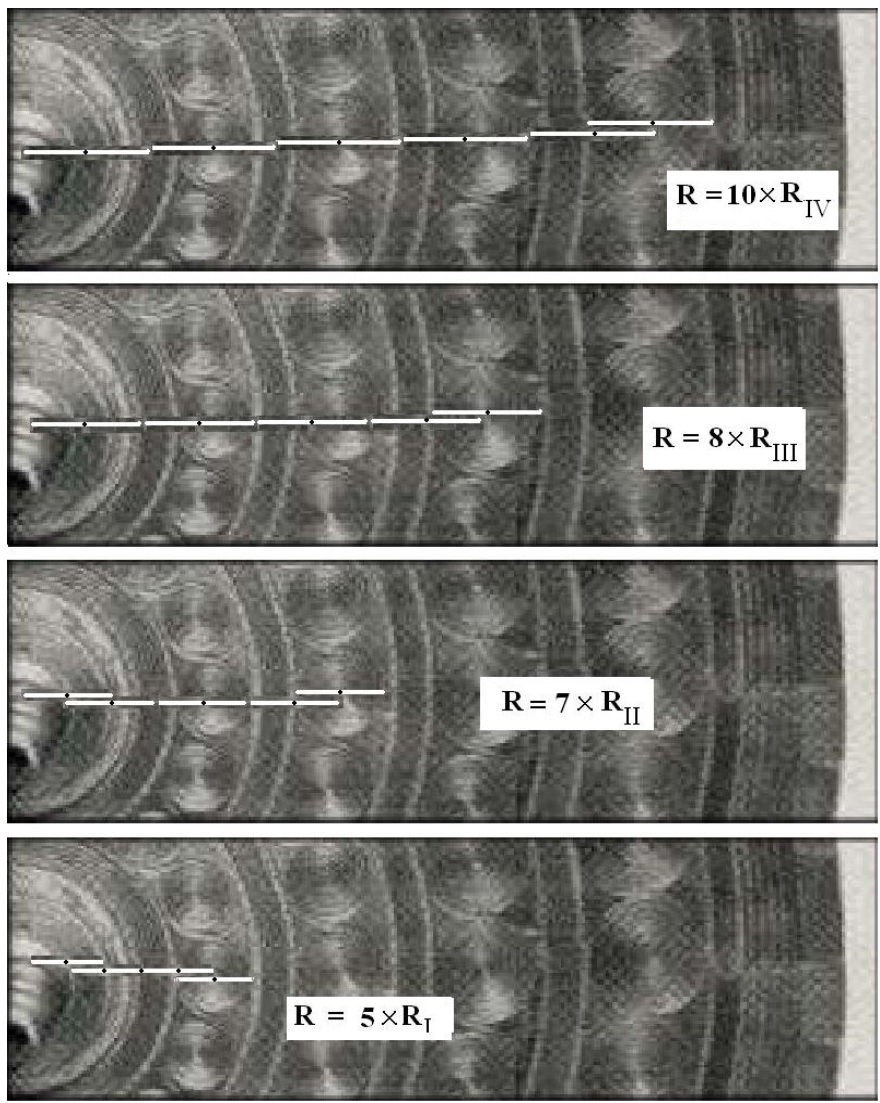

**Fig.3** - Probably the artist used a multiple of the spiral radius, taken on the radial direction, to determine the radius of the circumference which is the locus of the spiral centres, suitable for the decoration. We can obtain the multiple for each annulus directly from the picture.